\documentclass[graybox]{svmult}
\usepackage{mathptmx, helvet, courier, makeidx, graphicx, multicol, type1cm}
\usepackage{keyval,cite}
\usepackage[bottom]{footmisc}

\def\be {\begin{equation}}
\def\ee {\end{equation}}
\def\bea {\begin{eqnarray}}
\def\eea {\end{eqnarray}}
\def\bc {\begin{center}}
\def\ec {\end{center}}

\title*{Three-loop HTLpt thermodynamics at finite temperature and chemical potential}
\tocauthor{}
\author{Aritra Bandyopadhyay, Najmul Haque, Munshi G. Mustafa, Michael Strickland \and Nan Su}
\authorrunning{A. Bandyopadhyay et al.}
\institute{Aritra Bandyopadhyay, Munshi G. Mustafa \at Theory Division, Saha Institute of Nuclear Physics, 1/AF Bidhannagar, Kolkata 700064, India,\\ \email{aritra.bandyopadhyay@saha.ac.in, munshigolam.mustafa@saha.ac.in}
\and Najmul Haque, Michael Strickland \at Department of Physics, Kent State University, Kent, OH 44242, United States,\\ \email{nhaque@kent.edu, mstrick6@kent.edu}
\and Nan Su \at Institut f{\"u}r Theoretische Physik, Goethe-Universit{\"a}t Frankfurt am Main, D-60438 Frankfurt, Germany, \email{nansu@th.physik.uni-frankfurt.de}}
\date{\today} 

\begin{document}

\maketitle 

\abstract{
In this proceedings we present a state-of-the-art method of calculating thermodynamic potential at finite temperature and finite chemical potential, using Hard Thermal Loop perturbation theory (HTLpt) up to next-to-next-leading-order (NNLO). The resulting thermodynamic potential enables us to evaluate different thermodynamic quantities including pressure and various quark number susceptibilities (QNS). Comparison between our analytic results for those thermodynamic quantities with the available lattice data shows a good agreement. }

\abstract*{
In this proceedings we present a state-of-the-art method of calculating thermodynamic potential at finite temperature and finite chemical potential, using Hard Thermal Loop perturbation theory (HTLpt) up to next-to-next-leading-order (NNLO). The resulting thermodynamic potential enables us to evaluate different thermodynamic quantities including pressure and various quark number susceptibilities (QNS). Comparison between our analytic results for those thermodynamic quantities with the available lattice data shows a good agreement. }

\section{Introduction}
\label{sec:1}
With the increasing advancements of relativistic heavy ion colliders, the study of the color deconfined quark gluon plasma (QGP) phase has also become more extensive. From the last couple of decades, paired with the recurring successes of non-perturbative methods like lattice QCD (lQCD) simulations, one unavoidable question also haunted the theoreticians, now and again. How far can we trust those ideas which describes the QGP, based on perturbation theory? To answer this question, along with the high loop order calculations now one is also interested in the comparison of those analytic results with the relevant lattice QCD data.  

QCD pressure in perturbation theory for both zero and finite chemical potentials has already been calculated and results are known up to $\mathcal{O}(g^6\ln g)$~\cite{ipp, vuorinen1, vuorinen2}. The results show a poor convergence in the phenomenologically important region near the transition temperature. Search for the root of this problem eventually suggests some systematic reorganization of bare perturbation theory. One of them, in which we will focus on, is the Hard Thermal Loop perturbation theory (HTLpt), developed by Andersen, Braaten and Strickland~\cite{andersen1, andersen2} using the concept of Hard Thermal Loops, proposed by Braaten and Pisarski~\cite{pisarski2}, almost two decades ago. Since then, HTLpt has become one of the primary tools for evaluating thermodynamic quantities and it has already been used in the evaluation of QCD thermodynamic potentials up to one-loop (LO)~\cite{andersen3, sylvain1, purnendu1, najmul11, najmul12, najmul13}, two-loop (NLO)~\cite{andersen4, najmul2, najmul2qns} and recently even up to three-loop (NNLO)~\cite{3loopglue1, 3loopqcd1, 3loopqcd2, najmul3} for both zero and finite chemical potentials~\cite{3loopjhep, proceedings3loop}. 

In this proceedings, we present the recent evaluation of the NNLO QCD thermodynamic quantities \cite{najmul3,3loopjhep} in HTLpt at finite temperature and finite chemical potentials both for flavor-independent and flavor-dependent cases. We emphasize that this NNLO calculation of thermodynamic potential is completely analytic and gauge independent.

\section{Results}
\label{sec:2}
In the present section we show some plots showing the behavior of the thermodynamic quantities based on the final results from \cite{najmul3,3loopjhep}. Though we used both the one-loop and three-loop running coupling, but in this proceedings we show results only with the former one. By taking the value of $\alpha_s(1.5 $GeV$) = 0.326$ from lattice measurements~\cite{latticealpha}, the fixed value of the scale $\Lambda_{\overline{MS}}$ comes out to be $176$ MeV. For purely gluonic and fermionic Feynman diagrams, two separate renormalization scales, $\Lambda_{g}$ and $\Lambda$,  are used respectively. We use the debye mass prescription introduced by Braaten and Nieto in ~\cite{braatennieto2} which is generalized to finite chemical potential by Vuorinen in ~\cite{vuorinen1,vuorinen2}.

\begin{figure}
\begin{center}
\includegraphics[scale=0.6]{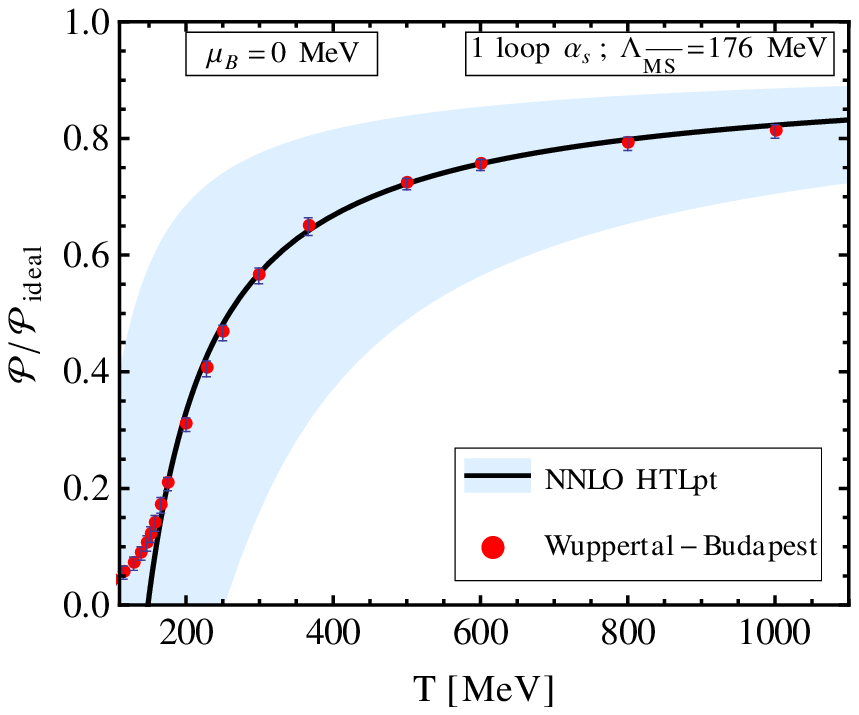}
\includegraphics[scale=0.6]{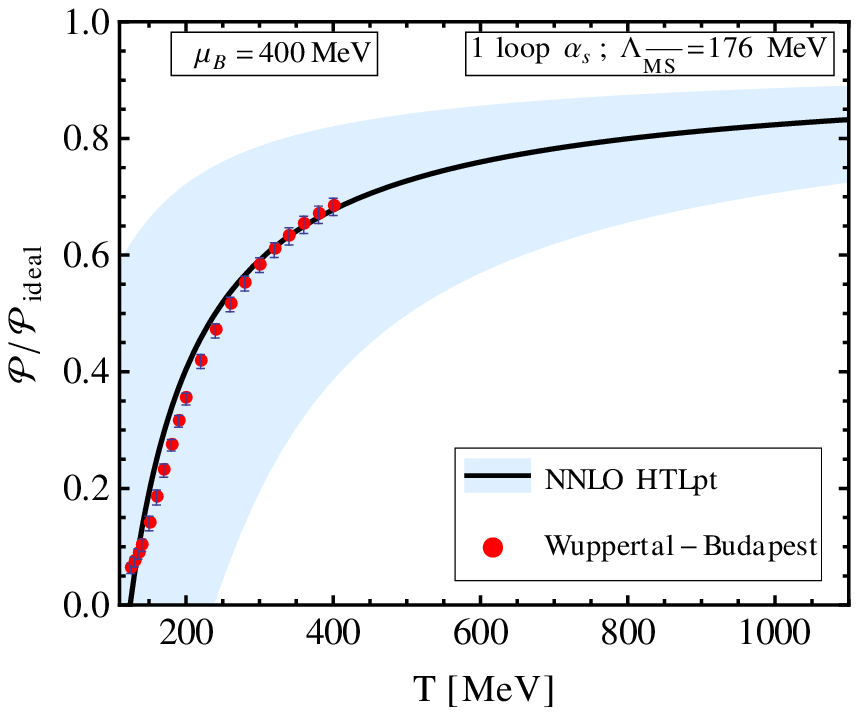}
\caption{Scaled NNLO HTLpt pressure for $N_f=2+1$ at $\mu_B = 0$ (left) and $\mu_B = 400$ MeV (right), compared with lattice data from
 Borsányi et al. \cite{borsanyi1,Borsanyi:2012uq}}
\label{fig:1}  
\end{center}
\end{figure}

In Fig.\ref{fig:1} we show the scaled NNLO HTLpt pressure for both $\mu_B = 0$ and $\mu_B = 400$ MeV, which can be obtained directly from the NNLO thermodynamic potential. In these plots along with the others to follow, the thick black line represents the result obtained using the central values of renormalization scales, e.g. $\Lambda_{g} = 2\pi T$ and $\Lambda = 2\pi\sqrt{T^2+\mu^2/\pi^2}$. The surrounding light-blue band specifies the variation of the result by varying these scales with a factor of two, e.g. $\pi T\le \Lambda_{g}\le 4\pi T$. The NNLO pressure is compared with available lQCD data which shows a very good agreement~\cite{borsanyi1,Borsanyi:2012uq}. 

\begin{figure}
\begin{center}
\includegraphics[scale=0.6]{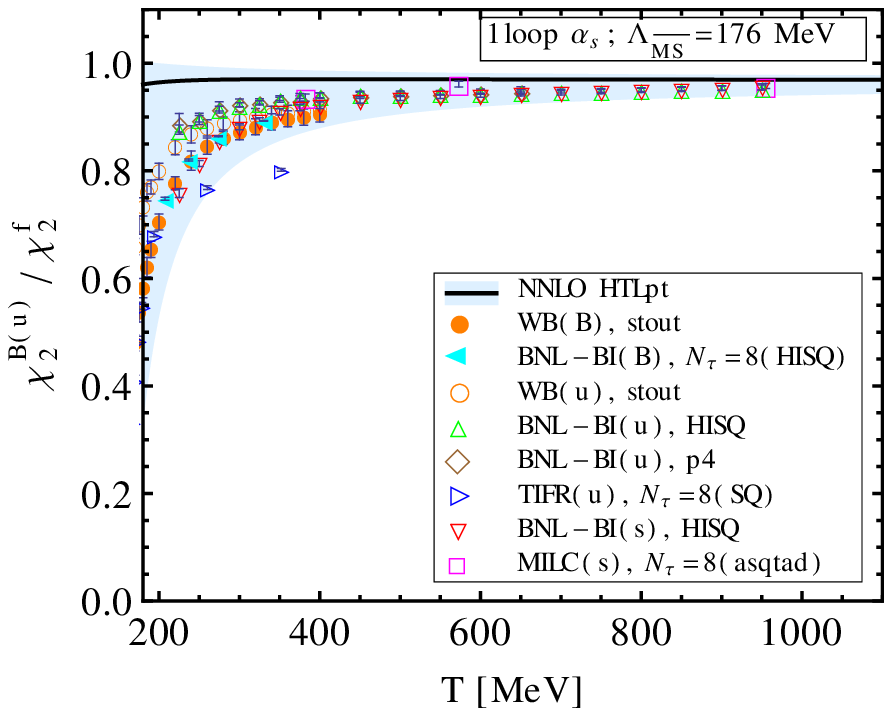}
\includegraphics[scale=0.6]{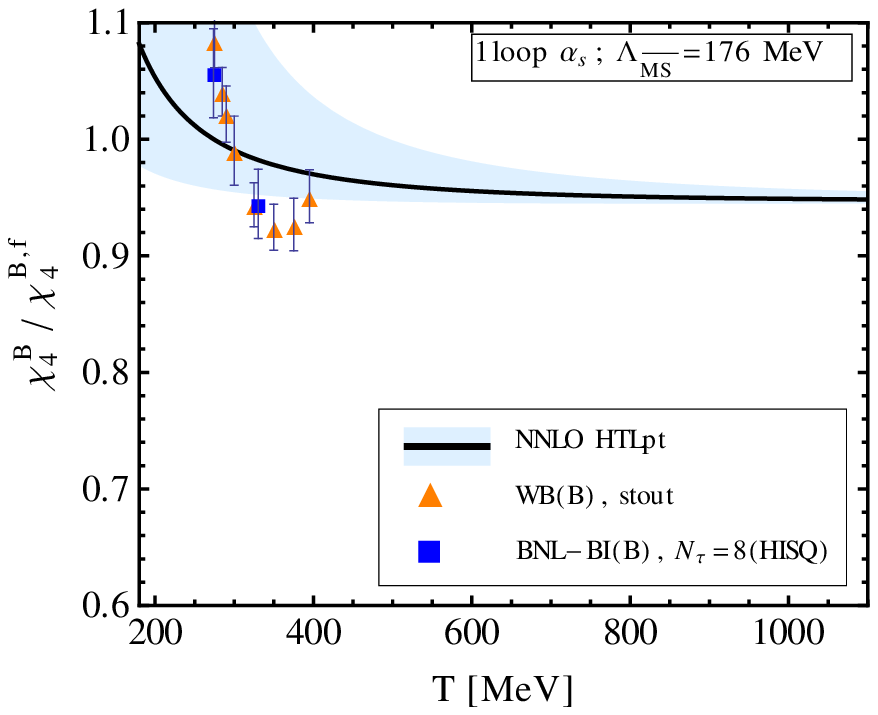}
\caption{For $N_f=2+1$, scaled NNLO HTLpt second order (left) and fourth order (right) baryon number susceptibility compared with various lattice data, labeled as WB, BNL-BI(B), BNL-BI(u,s), MILC, and TIFR come from Refs.~\cite{borsanyi2}, \cite{bnlb1}, \cite{bnlb2}, \cite{milc}, and \cite{TIFR}, respectively.}
\label{fig:2}  
\end{center}
\end{figure}

Usually the baryon number susceptibility (BNS) is related to the quark number susceptibilities~\cite{3loopjhep,najmul3}. If one assumes flavor independent unanimous quark chemical potential ($\mu_u=\mu_d=\mu_s=\mu$), then BNS can be calculated in a straightforward way by taking the derivatives of the pressure with respect to $\mu$. In Fig.\ref{fig:2} we compare the scaled second-order (left) and fourth-order (right) BNS with various lattice data~\cite{borsanyi2,bnlb1,bnlb2,milc,TIFR}.

\begin{figure}
\begin{center}
\includegraphics[scale=0.6]{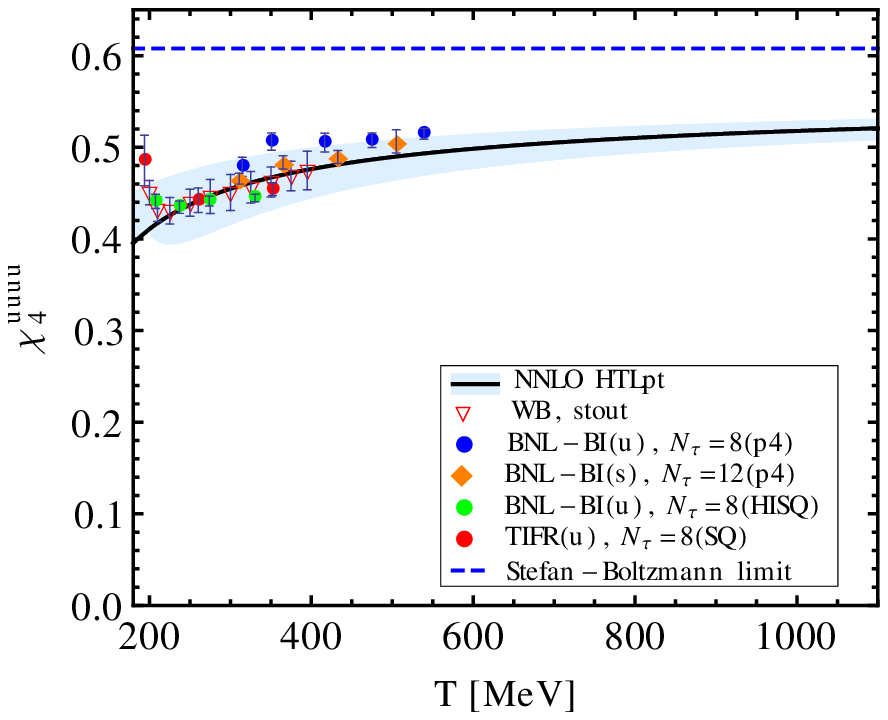}
\includegraphics[scale=0.6]{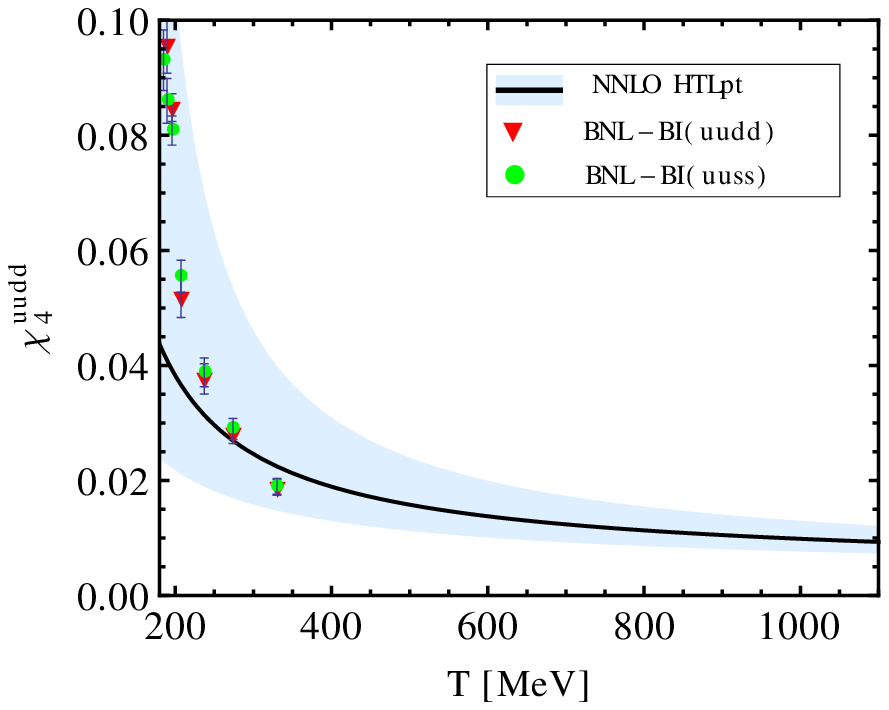}
\caption{For $N_f=2+1$, NNLO HTLpt fourth order diagonal single quark number susceptibility (left) and the only non-vanishing fourth order off-diagonal quark number susceptibility (right), compared with various lattice data, labeled as BNL-BI(uudd), BNL-BI(u, s), BNL-BI(uuss), and TIFR taken from Refs. \cite{bnlb1}, \cite{bnlb2}, \cite{bnlb3}, and \cite{TIFR}, respectively. In the left figure the dashed blue line indicates the Stefan-Boltzmann limit for this quantity.}
\label{fig:3}  
\end{center}
\end{figure}

To compute single quark number susceptibilities (sQNS), flavor dependent quark chemical potentials are taken into account in the general expression of NNLO thermodynamic potential~\cite{3loopjhep}. Derivatives with respect to $\mu_i$ yield either diagonal (same flavor on all indices) or off-diagonal (different flavor on some or all indices) susceptibilities.  In Fig.\ref{fig:3} we compare NNLO HTLpt fourth-order diagonal sQNS (left) and the only non-vanishing fourth-order off-diagonal sQNS (right) with available lQCD data from various groups~\cite{bnlb1,bnlb2,bnlb3,TIFR}. 

\section{Conclusions}

In this proceedings, the NNLO~QCD thermodynamic potential and various thermodynamic quantities within 3-loop HTLpt at finite temperature and chemical potential(s) are presented. Results obtained using the central values of the renormalization scales are well in accord with the accessible lattice data, for some of the quantities even down to phenomenologically more important temperatures relevant to LHC. On the ground of these perturbative results, one may further study other important facets of Quark Gluon Plasma, e.g. transport properties.

\end{document}